\documentstyle[12pt]{article}
\def\be{\begin{eqnarray}}
\def\ee{\end{eqnarray}}
\def\ba{\begin{array}}
\def\ea{\end{array}}

\begin{document}

\begin{center}
{\bf\LARGE {Parity Violation and Arrow of Time in \vskip 3mm
Generalized Quantum Dynamics}}
\end{center}

\vskip 1cm

\begin{center}
{\bf \large {Vadim V. Asadov$^{\star}$\footnote{asadov@neurok.ru}
and Oleg V.
Kechkin$^{+\,\star}$}\footnote{kechkin@depni.sinp.msu.ru}}
\end{center}

\vskip 5mm

\begin{center}
$^+$Institute of Nuclear Physics,\\
Lomonosov Moscow State University, \\
Vorob'jovy Gory, 119899 Moscow, Russia
\end{center}

\vskip 5mm

\begin{center}
$^\star$Neur\,OK--III,\\
Scientific park of MSU, Center for Informational Technologies--104,\\
Vorob'jovy Gory, 119899 Moscow, Russia
\end{center}

\vskip 1cm

\begin{abstract}
It is shown, that parity violation in quantum systems can be a
natural result of their dynamical evolution. The corresponding
(completely integrable) formalism is based on the use of quantum
theory with complex time and non-Hermitian Hamiltonian. It is
demonstrated, that starting with total symmetry between left and
right states at the initial time, one obtains strictly polarized
system at the time infinity. The increasing left-right asymmetry
detects a presence of well-defined arrow of time in evolution of the
system. We discuss possible application of the general formalism
developed to construction of modified irreversible dynamics of
massless Dirac fields (in framework of superstring theory, for
example).
\end{abstract}

\vskip 0.5cm

PACS No(s).\, : 05.30.-d,\,\,05.70.-a.

\vskip 1cm

The second low of thermodynamics states, that all closed dynamical
systems have essentially irreversible evolution. In this connection,
one says about `arrow of time',\, which separates initial and final
states of the system in absolute manner, and must be guaranteed by
any realistic dynamical theory without fail \cite{atf}--\cite{atl}.

Recently we have shown, that quantum theory with complex time
parameter ($\tau$), and non-Hermitian Hamiltonian structure ($\cal
H$) provides natural framework for quantum systems with irreversible
dynamics \cite{q-1}. In this generalized quantum theory, the state
vector $\Psi$ is analytic function on $\tau$, defined by the
conventional Schr\"{o}dinger's equation\be\label{G3} i\hbar
\Psi_{,\tau}={\cal{H}}\Psi.\ee Here, in the general case,
${\cal{H}}$ must be analytic on $\tau$ too, but we put
${\cal{H}}_{,\tau}=0$ in any concrete consideration. Then, in
\cite{q-1} also it was argued, that one can perform the
parametrization \be\label{G5} \tau=t-\,i\,\frac{\hbar}{2}\,\beta,
\quad {\cal{H}}=E-\,i\,\frac{\hbar}{2}\,\Gamma\ee for the quantities
$\tau$ and ${\cal{H}}$ in terms of the `usual'\, time $t$, the
inverse absolute temperature $\beta=1/kT$ (where $k$ is the Bolzman
constant), and also the energy and decay operators $E$ and $\Gamma$,
respectively. In the theory under consideration, we impose the
restriction \be\label{G4'}[E,\,\,\Gamma]=0, \ee i.e., we suppose,
that the decay operator $\Gamma$ must be taken in the form of a
symmetry operator of the system. The main statement is that the
average value $ \bar{\Gamma}=\Psi^+{{\Gamma}}\Psi/\Psi^+\Psi$ of the
decay operator is not increasing quantity for all physically
well-motivated temperature regimes $\beta=\beta(t)$ \cite{q-1}.
Below we prove this statement for the quantum systems with the
operator $\Gamma$ of the parity type and show, that the
corresponding dynamics possesses arrow of time in the transparent
form. Thus, we would like to demonstrate, that left-right asymmetry
in the quantum systems can be a natural result of their dynamical
evolution with arbitrary initial data taken. Moreover, we show how
the generalized quantum dynamics under consideration maps completely
left-right symmetric initial state to strictly polarized (left)
final one for the arbitrary temperature regime $\beta=\beta(t)$.

Thus, let us study a system, which consists of two parts -- its \,
`left' and `right' constituents -- in accordance to specification of
the symmetry operator $\Gamma$ taken. Namely, we define it
proportional to some parity operator; this leads to the double
eigenstate realization $\psi_{\pm n}$ for the all energy eigenvalues
$E_n$ . The corresponding eigenvalue problem reads: \be\label{5}
E\,\psi_{\pm}=E_n\,\psi_{\pm}, \qquad
\Gamma\,\psi_{\pm}=\mp\gamma\,\psi_{\pm}. \ee We choose the common
proper basis $\{\psi_{\pm}\}$ of the energy and decay operators in
the orthonormal form (from this it follows, that
$\psi^+_{\pm\,n_1}\psi_{\pm\,n_2}= \delta_{n_1n_2}$ and
$\psi^+_{+\,n_1}\psi_{- \,n_2}=0$). For definiteness, we name the
eigenstates $\psi_{+\,n}$ and $\psi_{-\, n}$ as the `left'\, and
`right'\,  ones, respectively, and take $\gamma>0$.

Then, performing standard calculations, one obtains the
probabilities $P_{\pm n}$ to find the system in its left and right
eigenstate realizations \cite{q-1}. The result reads: \be\label{6}
P_{\pm n}=\frac{w_{\pm n}}{Z}, \ee where \be\label{6'}w_{\pm
n}=e^{\pm \gamma t}\,\tilde w_{\pm n},\qquad Z=e^{\gamma t}\tilde
Z_{+}+ e^{-\gamma t}\tilde Z_{-},\ee and also \be\label{7} \tilde
w_{\pm n}=\rho_{\pm n}e^{-E_n\beta}, \qquad \tilde
Z_{\pm}=\sum_n\tilde w_{\pm n}, \ee where $\rho_{\pm n}=|C_{\pm
n}|^2$. Our goal is to study evolution of the left and right parts
of the system, which we relate to the subsets $\{P_{+\,n}\}$ and
$\{P_{-\,n}\}$ of the basis probabilities.

It is not difficult to prove, that result of this evolution can be
expressed in terms of the effective `left\, and `right'\,
probabilities \be\label{8} \tilde p_{\pm\, n}=\frac{\tilde w_{\pm \,
n}}{\tilde Z}, \ee which have standard pure $\beta$-dependent form
in view of Eq. (\ref{7}). Actually, in the isothermal case (with
$\beta={\rm const}$), one obtains immediately, that \be\label{9}
{\rm lim}_{_{t\rightarrow \pm \infty}}\,P_{\pm\, n}\left ( t\right
)=\tilde p_{\pm\,n}, \qquad {\rm lim}_{_{t\rightarrow  \pm
\infty}}\,P_{\mp\, n}\left ( t \right )=0.\ee Thus, in the
isothermal evolution, this quantum system transforms from its left
realization to the right one. From this it follows, that
\be\label{10} {\rm lim}_{_{t\rightarrow  \pm \infty}}\bar\Gamma\left
( t\right )=\pm\gamma,\ee i.e. that the total shift of average value
of the decay operator $\Gamma$ is equal to $2\gamma$. It is clear,
that the same situation takes place for the arbitrary thermodynamic
regime, which admits the asymptotic inverse temperatures
$\beta_{\pm}={\rm lim}_{_{t\rightarrow \pm\infty}}\beta(t)$. In this
case, one must replace $\tilde p_{_{\pm \, n}}$ by $\tilde p_{_{\pm
\, n}}(\beta_{\pm})$ in Eq. (\ref{9}).

Now let us consider an intriguing special situation, where the
quantum system reaches a total symmetry between its left and right
constituents at some finite time $t_{\star}$. Namely, we are
interested in the dynamics with $P_{+\,n}( t_{\star})= P_{-\,n} (
t_{\star})$ for all values of the collective index $n$. Putting
$t_{\star}=0$ (without loss of generality), one obtains the system
with $\rho_{+\,n}=\rho_{-\,n}\equiv \rho_{n}$, and \be\label{11}
P_{\pm\,n}=\frac{\tilde p_n}{1+e^{\mp2\gamma t}}, \ee where $\tilde
p_n=\tilde w_n/\tilde Z$,\, $\tilde Z=\sum _n\tilde w_n$, and
$\tilde w_n=\rho_ne^{-E_n\beta}$. Thus, in this case one deals with
the single effective system related to the set of the
$\beta$-dependent probabilities $\tilde p_n$. For the quantum system
under consideration, the average value of the decay operator reads:
\be\label{12} \bar\Gamma=\gamma\,\tanh\, \gamma t. \ee Thus, the
quantity $\bar\Gamma (t)$ is a monotonously increasing Lyapunov
function for the attractor left state. It detects a dynamical parity
breaking effect for the system. This function defines universal
arrow of time: its form does not depend on the concrete temperature
regime $\beta=\beta (t)$, and demonstrates fundamentally
irreversible character of the dynamics under consideration.

One important realization of the generalized quantum formalism
developed above is related to the theory of massless Dirac field.
Namely, the spinor system with the standard (massless) energy
operator $E=\vec\alpha\,\vec p$ \,(where
$\alpha_k=\gamma_0\gamma_k$, and $\vec p$ \, is the momentum), and
the decay operator $\Gamma=\gamma\,\gamma_{5}$ of the parity type
(here $\gamma={\rm const}$), satisfy the restriction (3) and belongs
to the theory class considered. Thus, in complete agreement with the
general results presented above, the originally mixed left-right
massless Dirac quantum system transforms to the strictly polarized
asymptotic form in the case of $\gamma\neq 0$. One can choose a sign
of this free constant parameter in appropriate way to obtain a
`visible'\, helicity of the final state, and to identify it with the
real neutrino system.

Actually, it is a well known fact, that the real neutrinos are
`left'\, (whereas the antineutrinos are `right'),\, and that this
circumstance seems intriguing in view of its `pure random'\, status
in the standard particle physics
\cite{PartPhysCosm-f}-\cite{PartPhysCosm-l}. We think, that the new
approach, which provides the dynamical solution for helicity
asymmetry problem, is more natural than the fundamentally asymmetric
standard scheme. The new approach relates parity violation in the
real neutrino system with arrow of time in its evolution (and with
action of the second law of thermodynamics in the Universe after
all). Thus, in the framework of generalized quantum theory, this
random feature of the real physical world becomes a consequence of
its irreversible history from the universal thermodynamical point of
view.

We would like to stress, that our modification of the massless
spinor system can be used for the supergravity and string theory
generalizations to guarantee the presence of arrow of time in these
dynamical systems. It seems especially important in cosmological and
black hole physics context, where the second law of thermodynamics
must be taken into account without any doubt.

\vskip 10mm \noindent {\large \bf Acknowledgements}

\vskip 3mm \noindent We would like to thank prof. B.S. Ishkhanov for
many important discussions and private talks which were really
useful for us during this work preparation. One of the authors (Oleg
V. Kechkin) was supported by grant ${\rm MD \,\, 3623.\, 2006.\,
2}$.

\end{document}